\documentclass[twocolumn]{aastex631} 

\usepackage[]{mathtools}
\usepackage{bm}
\usepackage{graphicx}
\usepackage{subfigure} 
\usepackage{float}
\usepackage{verbatim}
\usepackage{booktabs}
\usepackage{multirow}
\usepackage{color}
\usepackage{mathtools}

\begin{document}

\title{Rapid pre-merger localization of binary neutron stars in third generation gravitational wave detectors}

\author[0000-0002-3033-6491]{Qian Hu}
\email{q.hu.2@research.gla.ac.uk}
\author[0000-0002-6508-0713]{John Veitch}
\email{John.Veitch@glasgow.ac.uk}
\affiliation{Institute for Gravitational Research, School of Physics and Astronomy, University of Glasgow, Glasgow, G12 8QQ, United Kingdom}

\begin{abstract} 
Pre-merger localization of binary neutron stars (BNSs) is one of the most important scientific goals for the third generation (3G) gravitational wave (GW) detectors.
It will enable the electromagnetic observation of the whole 
process of BNS coalescence, {especially for the pre-merger and merger phases which have not been observed yet, opening a window for deeper understandings of compact objects.} 
To reach this goal, we describe a novel combination of multi-band matched filtering and semi-analytical localization algorithms to achieve early-warning localization of long BNS signals in 3G detectors. Using our method we are able to efficiently simulate one month of observations with a three-detector 3G network, and show that it is possible to provide accurate sky localizations more than 30 minutes before the merger.
Our simulation shows that there could be $\sim 10$ ($\sim 100$) BNS events
localized within 100~deg$^2$, 20 (6) minutes before merger, per month of observation. 
\end{abstract}

\section{\label{sec:intro} Introduction}
Since the first direct gravitational wave (GW) detection of the coalescence of binary neutron star (BNS) GW170817~\citep{abbott2017_GW170817ObservationGravitational} and its electromagnetic (EM) counterparts~\citep{abbott2017_MultimessengerObservationsBinary}, multi-messenger observation {of coalescing compact binaries} has become an
{important tool for astrophysics.}
Joint GW-EM observations can provide a comprehensive understanding of the formation and evolution of BNS, and {shed lights on physics around compact objects}~\citep{abbott2019_PropertiesBinaryNeutron,De:2018uhw,Mooley:2018qfh,Capano:2019eae,Nicholl:2017ahq, LIGOScientific:2018cki, Annala:2017llu, Kasen:2017sxr,Margalit:2017dij,Cowperthwaite:2017dyu, DES:2017kbs}. 

Rapid GW detection and accurate localization is key to joint GW-EM observations, as most EM facilities need the direction from the GW observation. In addition to capturing the afterglow of BNS coalescences, early EM observations could offer unique insights of phenomena in BNS that happen prior to or near the merger {(e.g. tidal disruptions, magnetosphere interactions, r-process nucleosynthesis)} and help build a picture of the entire evolution of kilonova in multiple frequency bands~\citep{Cooper:2022slk,Most:2020ami,Nicholl:2017ahq,Metzger:2013cha,Radice:2018pdn,Metzger:2019zeh}. Early detection and localization of BNS are therefore of great importance in GW astronomy. It has previously been demonstrated that there is a non-zero probability of detecting and localizing pre-merger BNS events with {current}~\citep{Kovalam:2021bgg,Magee:2021xdx} and {near-future}~\citep{Sachdev:2020lfd,Nitz:2020vym,Magee:2022kkc,Banerjee:2022gkv} GW observatories, typically within seconds to one minute before merger. {\citet{Chaudhary:2023vec} has recently investigated early warning for the fourth observing run of LIGO-Virgo-KAGRA collaboration, with multiple detection pipelines already equipped for early warning searches~\citep{Chu:2020pjv, Nitz:2018rgo, Messick:2016aqy, Aubin:2020goo}.} {Machine learning based methods for early warning detection are also making rapid progresses~\citep{Baltus:2021nme, Wei:2020sfz, Yu:2021vvm, baltus2021detecting} }


Being limited by low sensitivities below 20\,Hz, the BNS signal is only detectable for $\sim$1 minute in current GW detectors. {Given the communication time delay between multi-messenger community and the $\sim 10-100$\,seconds slew time of modern telescopes~\citep{Banerjee:2022gkv}, it is basically impossible to capture pre-merger or near-merger transients from BNS coalescence} {without fore-warning.}
Several third-generation (3G) GW detectors have been proposed, including Einstein Telescope (ET)~\citep{Punturo:2010zz} and Cosmic Explorer (CE)~\citep{reitze2019_CosmicExplorerContribution,evans2023cosmic}, with low frequency sensitivities significantly improved. These would allow us to detect BNS signals more than 30 minutes before the merger, rendering precise early warning localization possible~\citep{Nitz:2021pbr,Branchesi:2023mws,Borhanian:2022czq,ronchini2022_PerspectivesMultimessengerAstronomy,chan2018_BinaryNeutronStar, Akcay:2018aqh}. 

\begin{figure}
    \includegraphics[width=0.46\textwidth]{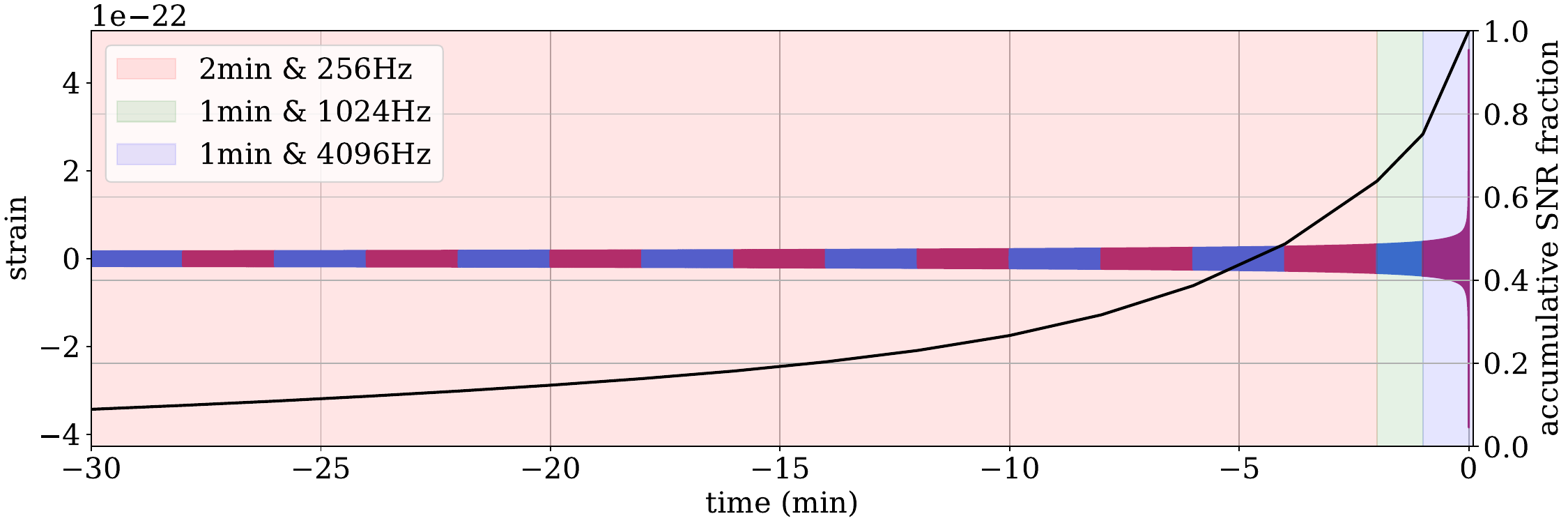}
    \centering
    \caption{\label{fig:td_waveform_smooth} Multi-banding scheme for this work. \textbf{Left axis:} An illustration of a chopped GW waveform that is alternately colored for different waveform bins and sampling frequencies. We use two-minute segments with 256\,Hz sampling frequency for waveforms from 60 minutes before merger to two minutes before merger, and one-minute segments for the last two minutes with 1024\,Hz and 4096\,Hz sampling frequencies, respectively. \textbf{Right axis:} The cumulative SNR of the signal, showing the contribution of SNR from each time segment.}
\end{figure}

However, data analysis of BNS in 3G detectors can be challenging. {The long signal makes matched filtering extremely expensive to perform, and it is modulated by changes of the antenna response functions due to the Earth's rotation. Neglecting the Earth's rotation should have little impact on detection, as signal-to-noise ratio (SNR) is mostly contributed by the last stages of the signal (see Fig.~\ref{fig:td_waveform_smooth}). However, ignoring it causes loss of information and could lead to biases in parameter estimation, especially for sky location parameters.}
In this work, we demonstrate that multi-band analysis~\citep{Aubin:2020goo,Adams:2015ulm,Cannon:2011vi} is an effective way of solving these issues, and fast localization algorithms can be built upon multi-band detection statistics. 

Multi-band analysis is based on the fact that orbital evolution of the {quasi-circular} BNS inspiral stage is well modeled. Observable BNSs are not likely to have large spins~\citep{Abbott2020b,abbott2019_PropertiesBinaryNeutron,abbott2017_GW170817ObservationGravitational} or precession, and the frequency evolves as 
\begin{equation}
    \label{eq:foftau}
    f(\tau) = 134~\mathrm{Hz} \left(\frac{1.21M_\odot}{\mathcal{M}_c}\right)^{5/8} \left(\frac{1~\mathrm{s}}{\tau}\right)^{3/8} 
\end{equation}
to the Newtonian order~\citep{GWbook1}, where $f$ is GW frequency, $\tau$ is the time before merger, and $\mathcal{M}_c$ is the chirp mass of the binary. {The monotonic evolution of GW frequency ensures a one-to-one correspondence between time segments and frequency bands, i.e.,} chopping the GW waveform into multiple time segments, $[t_n, t_{n-1}), [t_{n-1}, t_{n-2}), \dots, [t_{1}, t_{0})$, results in a corresponding sequence of frequency bands $[f_n, f_{n-1}), [f_{n-1}, f_{n-2}), \dots, [f_{1}, f_{0})$ defined via Eq.~\ref{eq:foftau}. One can choose the length of time intervals such that within each interval the Earth's rotation can be ignored, i.e. the detectors' antenna response functions can be assumed constant and current matched filtering techniques can be directly employed.
One can also down-sample the data in each frequency band according to its highest frequency to reduce computational cost. Results from each time segment can be combined in succession as new data comes in.

Fig.~\ref{fig:td_waveform_smooth} shows the multi-band scheme in this work. We consider the negative latency up to 60\,min, and choose 2\,min segments and 256\,Hz sampling frequency until the final two minutes. In the last two minutes the SNR grows rapidly, and the GW reaches high frequencies, therefore a finer time resolution is used to improve the detection and localization. {In addition to the limit from Nyquist–Shannon sampling theorem,  in practice we find that a sampling rate that is higher than Nyquist frequency could be helpful in localizing high SNR events.} As a demonstration, we equally divide the last two minutes, and employ sampling frequencies of 1024\,Hz and 4096\,Hz, respectively. {A more elaborate segmentation is also sensible {e.g. \citep{Morisaki:2021ngj}}, but we leave a comprehensive investigation of the multi-band scheme to future works.}

We will build a fast localization algorithm based on the above multi-banding scheme. Given the large number of BNS detections in the 3G detectors (typically $\sim 10^5$ per year ~\citep{Branchesi:2023mws,Borhanian:2022czq}), an efficient and light algorithm is necessary. While machine learning methods~\citep{gabbard2022_BayesianParameterEstimation,dax2021_RealtimeGravitationalwaveScience,Chatterjee:2022dik,Chatterjee:2022ggk} are gradually making progresses, \texttt{Bayestar}~\citep{Singer:2015ema}, which performs a five-fold numerical marginalization over nuisance extrinsic parameters, has been used as the standard low-latency localization approach {throughout the observing runs of the current generation of GW detectors}. It takes \texttt{Bayestar} $\sim 2$\,s to generate a skymap with 32 threads~\citep{Singer:2015ema}, and lower latency can be achieved with more CPUs {or narrower bandwidth. For instance, \citet{Magee:2021xdx} demonstrated $\sim 0.5$\,s of computation time of \texttt{Bayestar} with sufficient ($> 100$) threads for early warning triggers, and $\sim 1.1$\,s for full bandwidth triggers.} In this work, to reduce the computational cost of dealing with vast number of events, we will make use of a semi-analytical localization algorithm for GWs (\texttt{SealGW})~\citep{Hu:2021nvy} that has recently been implemented in the SPIIR detection pipeline~\citep{Chu:2020pjv} and is publicly available\footnote{\url{https://git.ligo.org/spiir-group/SealGW}}. \texttt{SealGW} performs a semi-analytical marginalization over nuisance parameters, achieves a faster performance {than Bayestar} and retains a reasonable accuracy. A more detailed description will be given in the following sections.

\section{\label{sec:sealgw} Localization for long signals}
As intrinsic parameters of GWs are initially estimated by matched filtering searches and their errors are semi-independent with errors in sky localization~\citep{Singer:2015ema}, they are treated as perfectly known in online fast localization. The Bayesian posterior probability distribution for extrinsic parameters can be written as 
\begin{equation}
    p(\boldsymbol{\vartheta_{\mathrm{ex}}} | d, \boldsymbol{\vartheta_{\mathrm{in}}}) \propto p(\boldsymbol{\vartheta_{\mathrm{ex}}}) p(d | \boldsymbol{\vartheta_{\mathrm{ex}}}, \boldsymbol{\vartheta_{\mathrm{in}}} ),
\end{equation}
where $d$ is the data, $p(\boldsymbol{\vartheta_{\mathrm{ex}}})$ is the prior distribution and $p(d | \boldsymbol{\vartheta_{\mathrm{ex}}}, \boldsymbol{\vartheta_{\mathrm{in}}} )$ is the likelihood. Among $\boldsymbol{\vartheta_{\mathrm{ex}}} = \{\alpha, \delta, t_c, \psi, r, \phi_c, \iota \}$, right ascension $\alpha$ and declination $\delta$ describes the source sky direction while the other nuisance parameters should be marginalized. Direct analytical marginalization is impossible, but after a parameter conversion 
\begin{equation}
    \boldsymbol{\vartheta_{\mathrm{ex}}} = \{\alpha, \delta, t_c, \psi, r, \phi_c, \iota \} \rightarrow \boldsymbol{\vartheta'_{\mathrm{ex}}} = \{\alpha, \delta, t_c, \boldsymbol{\mathrm{A}} \},
\end{equation}
where 
\begin{equation}
    \label{Amatrix}
    \mathbf{A} = \begin{pmatrix} \cos2 \psi & \sin 2\psi \\ -\sin2 \psi & \cos2\psi \end{pmatrix} \begin{pmatrix} \frac{1+\cos^{2}\iota}{2r} &  \\  & \frac{\cos\iota}{r} \end{pmatrix} \begin{pmatrix} \cos\phi_{c} & \sin\phi_{c} \\ -\sin\phi_{c} & \cos\phi_{c}. \end{pmatrix},
\end{equation}
the likelihood function appears as a Gaussian in the matrix $\boldsymbol{\mathrm{A}}$
and is therefore analytically tractable~\citep{Hu:2021nvy}. The posterior of the source sky location takes the form 
\begin{equation}
    p(\alpha, \delta | d, \boldsymbol{\vartheta_{\mathrm{in}}})=\int dt_c d^4\boldsymbol{\mathrm{A}} p(\boldsymbol{\vartheta'_{\mathrm{ex}}} | d, \boldsymbol{\vartheta_{\mathrm{in}}})  =  \int dt_c I(\rho(t_c), \alpha, \delta),
\end{equation}
where $\rho(t_c)$ is the SNR timeseries. The analytical expression of $I(\rho(t_c), \alpha, \delta)$, including a comprehensive introduction and tests of \texttt{SealGW}, can be found in \citet{Hu:2021nvy}. 

{Filtering each timeseries in the multi-band scheme produces a separate complex SNR time-series that must be coherently combined to achieve a precise localization~\citep{allen2012_FINDCHIRPAlgorithmDetection,Singer:2015ema}}. 
The gain in precision for long signals comes not only from the accumulation of absolute SNR, but also from phase drifts of the SNR timeseries due to the Earth's rotation because the rotation induces a longer equivalent network baseline~\citep{Wen:2010cr,Zhao:2017cbb,Baral:2023xst}.  
A direct combination (linear addition) of multiple SNR timeseries {is feasible for short signals, as it only} requires a set of combination parameters to align the template bands in time and phase~\citep{Adams:2015ulm}.
{However, for long signals, the direct combination scheme is no longer coherent because of the phase drifts due to the Earth's rotation. Including the phase drifts in combination parameters can lead to a coherent addition, but it loses information contained in the changing time delays between detectors which depends on the sky location.}
Therefore, instead of directly adding SNR, we multiply likelihoods from every band before marginalization over nuisance extrinsic parameters. Since {there is little overlap between bands (contributed by noise correlation and template overlaps which are windowed out)}, they can be treated as independent measurements, as if there are different detectors at different frequency bands.

\section{\label{sec:catalog} Catalog simulation}
To assess the performance of the above localization scheme, we simulate a mock BNS catalog, assuming a three-detector network with one triangle ET at Virgo site and two L-shaped CEs at LIGO Hanford and Livingston site, respectively. We use an analytical astrophysical population~\citep{oguri2018_EffectGravitationalLensing}
\begin{equation}
    R_{\mathrm{obs}}(z)=\frac{a_1 e^{a_2 z}}{e^{a_3 z}+a_4} \frac{1}{1+z}\frac{dV_\mathrm{c}}{dz} \mathrm{Gpc^{-3}yr^{-1}},
\end{equation}
Here $V_\mathrm{c}$ is the comoving volume and we employ Planck15 cosmology~\citep{Planck:2015fie}. $a_{\{2,3,4\}}$ are set to be $\{1.6, 2.1, 30 \}$ to model a peak at $z\sim 2$. $a_1$ is scaled to match the local BNS merger rate given by \citet{LIGOScientific:2020kqk} ($\mathcal{R}_{\mathrm{obs}}(z=0)=320_{-240}^{+490} \mathrm{Gpc^{-3}yr^{-1}}$). {We use the estimated median value of the merger rate in this work.} We simulate 68000 BNS sources within $z=3$, which corresponds to roughly one month of observations. We assume neutron star mass is uniformly distributed in $[1.1M_\odot, 2M_\odot]$ in the source frame and isotropic sky distribution and inclination. Note that the location and configuration of detector networks are not settled yet and subject to change, and {the BNS mass distribution and} the current merger rate density estimate have large uncertainties due to the as yet small number of BNS detections.

Signals are injected into Gaussian noise realizations and analyzed individually, i.e., we do not consider them to be overlapping with each other. Overlapping signals could cause dominant biases in parameter estimation, but mainly in the case when the merger times are very close~\citep{himemoto2021_ImpactsOverlappingGravitationalwave,pizzati2022_InferenceOverlappingGravitational,relton2021_ParameterEstimationBias,relton2022_AddressingChallengesDetecting,samajdar2021_BiasesParameterEstimation,Hu:2022bji}. {Among the 68000 BNS events evenly distributed in one month, roughly 1.3\% of them have another event ending $<0.5\,\mathrm{s}$ afterwards. Many of the signals are not actually detectable, further reducing the chance of significant bias. Even though the number of overlapping signals  could be large given the large number of events expected to be detected with 3G detectors,} {our simulation will still apply to the vast majority of non-interfering signals.}


We use the waveform model \texttt{TaylorF2}~\citep{Blanchet:1995ez, Poisson:1997ha, Buonanno:2009zt} to generate GW signals and map frequency to time before merger via the stationary phase approximation with Eq.~\ref{eq:foftau}. The 3.5 post-Newtonian waveform is a reasonable choice for analyzing quasi-circular inspiralling compact binaries~\citep{Faye:2012we}. Several works suggest some non-quasi-circular binaries, like precessing and eccentric systems, or systems with strong higher order emission, can be better localized~\citep{McIsaac:2023ijd,Tsutsui:2020bem,ma2017_GravitationalWaveSource,Kapadia:2020kss,Singh:2020lwx}. However, that would require novel search algorithms (e.g. ~\citet{Fairhurst:2019vut}) upon which new fast localization methods would have to be built, because current localization methods, including \texttt{Bayestar} and \texttt{SealGW}, are based on aligned-spinning waveform templates in which plus and cross polarizations of GWs only have a phase difference. 

We perform matched filtering assuming a perfect knowledge of intrinsic parameters and set total SNR$>$12 as the detection criterion, where total SNR is converted from {the multi-band matched filtering outputs} by the analytical expression of SNR at Newtonian order~\citep{Cannon:2011vi}. Matched filtering with known injection parameters is the ideal case, while in a realistic scenario one should build a template bank that achieves a reasonable match (e.g.~$>$97\%) everywhere in the parameter space. The purpose of this work is to assess the performance of the multi-band localization scheme. We leave a dedicated long signal early warning pipeline and simulations with more realistic mock data to future work.

\section{\label{sec:results} Results}
\begin{figure}
    \includegraphics[width=0.46\textwidth]{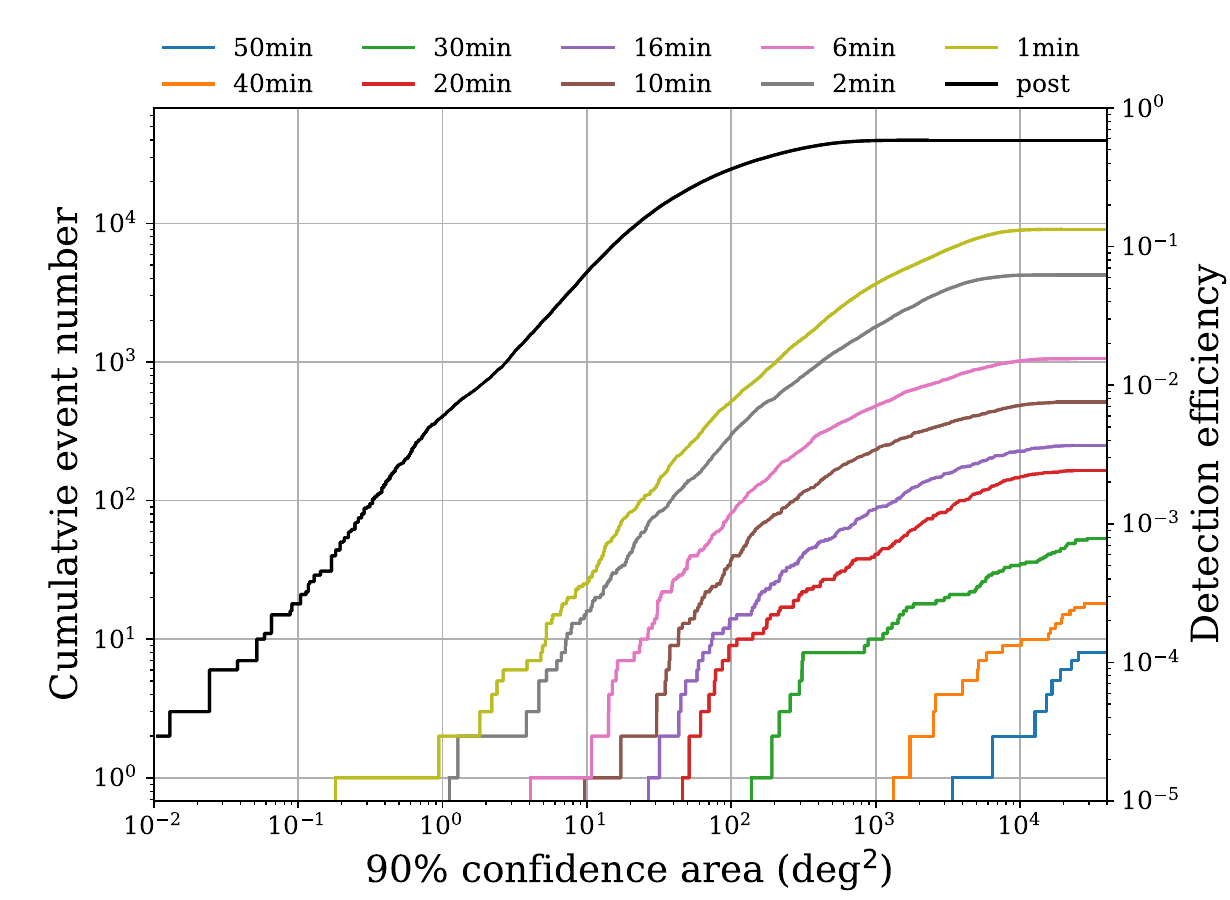}
    \centering
    \caption{\label{fig:areas90} Cumulative {number of detections and} 90\% confidence sky localization areas for the 68000 BNS simulations (roughly one month of observation). {The corresponding detection efficiency is labeled in the right $y$ axis.} We choose 10 different negative latencies (from 50 minutes to post-merger) and the curves show the cumulative distribution of 90\% areas of events that are detected at those times.}
\end{figure}

For each simulation, we perform multi-band matched filtering from 60 minutes before merger with low frequency cutoff at 5\,Hz. Fig.~\ref{fig:areas90} shows the cumulative number of events for different negative latencies. $\sim 10$ events can be localized within 100 deg$^2$ 20 minutes before merger, and 6 minutes before merger the number of events increases to $\sim 100$. {Also, $\sim 1-10$ events can be localized  within 10 deg$^2$ up to 6 minutes before merger. }

\begin{figure*}
    \centering
    \includegraphics[width=1\textwidth]{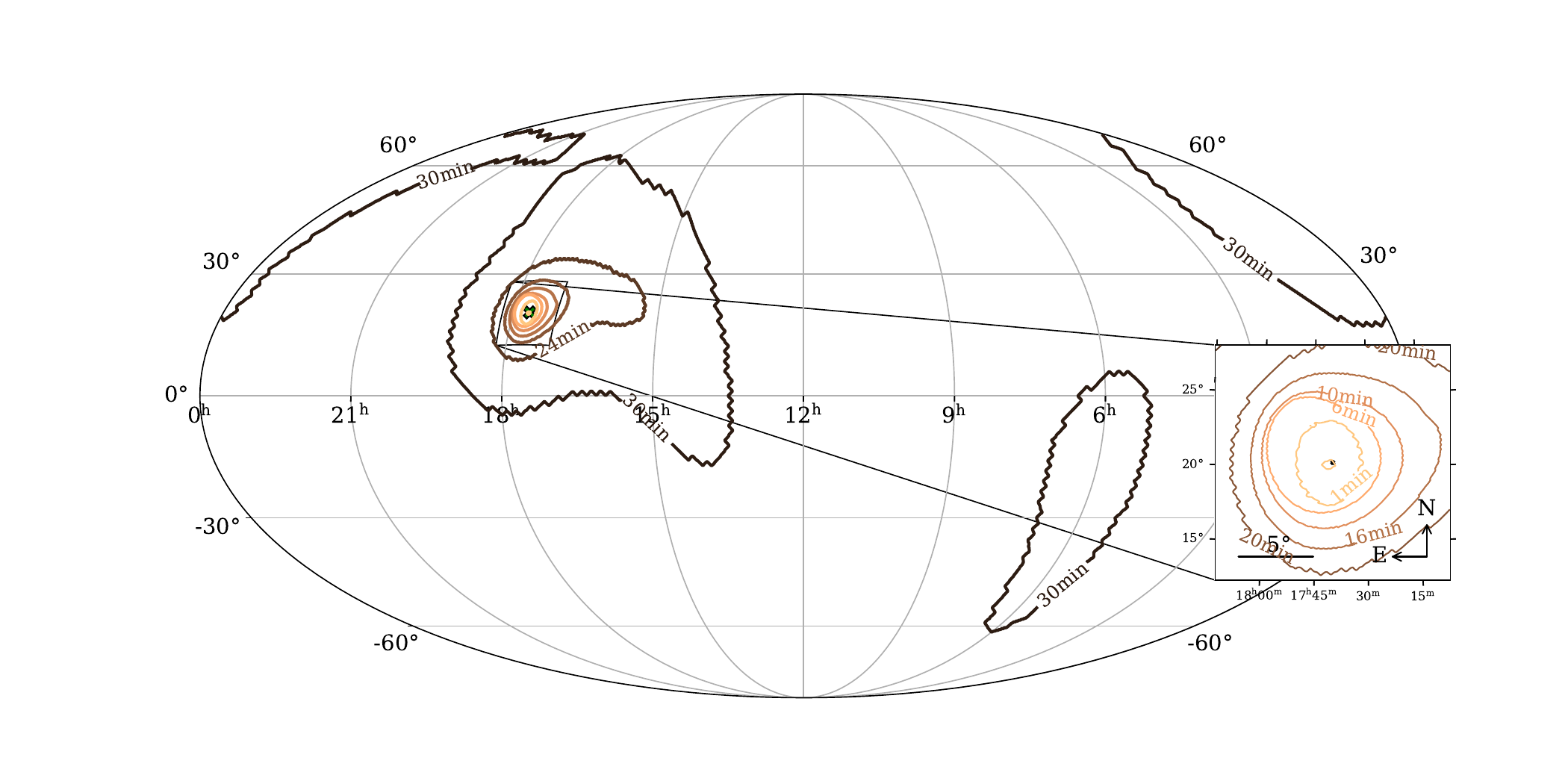}
    \includegraphics[width=1\textwidth]{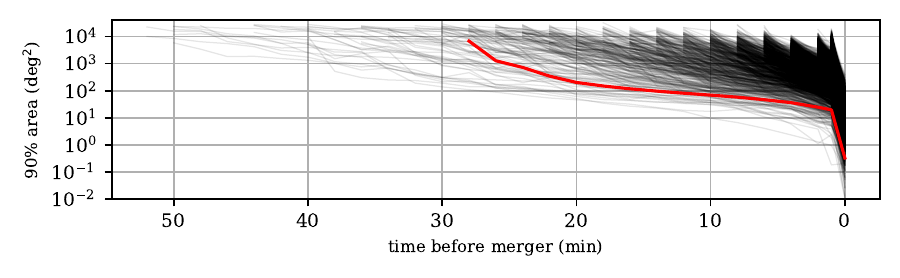}
    
    \caption{\label{fig:areasevo} Skymap evolution. \textbf{Upper panel}: An example skymap for a $1.4+1.4 M_\odot$ BNS at 1000 Mpc detected 30 minutes before merger {with a network SNR of 12. The SNR increases to 17 at 20 minutes before merger, 31 at 10 minutes, 95 at one minute and 130 after merger}. We show the 90\% localization contours at different negative latencies. The injection sky location is marked with a cross. \textbf{Lower panel}: Evolution of 90\% confidence localization areas of early warning events in our simulation. The example in upper panel is plotted in red line. }
\end{figure*}

Extreme early warnings are possible. Several events in our simulation are detected and {preliminarily localized} 40-50 minutes before merger and this number could be underestimated since our analysis has hard cutoffs at 5\,Hz and one-hour negative latency. ET would be able to collect sensible data down to $\sim3\,$Hz and trigger even earlier detections~\citep{Branchesi:2023mws,Borhanian:2022czq}. However, BNS with high negative latencies are not likely to be well localized until more data comes in, bringing higher SNRs, wider frequency bands and a longer equivalent network baseline. Multi-band analysis helps update detection statistics and skymaps on-the-fly. Fig.~\ref{fig:areasevo} shows the evolution of skymaps and localization areas in our simulation. The example skymap is from a $1.4+1.4 M_\odot$ BNS at 1000\,Mpc detected 30 minutes before merger. It presents nested contours with new bands combined in succession and is finally pinpointed within 0.2 deg$^2$, but is already well localized $\sim$10 minutes before the merger. The localization area traces in the lower panel show the decreasing rate of localization areas: those localized within 100 deg$^2$ $\sim 20$ minutes before merger in Fig.~\ref{fig:areas90} are generally detected 40-50 minutes before merger. 

\begin{figure}[t!]
    \includegraphics[width=0.46\textwidth]{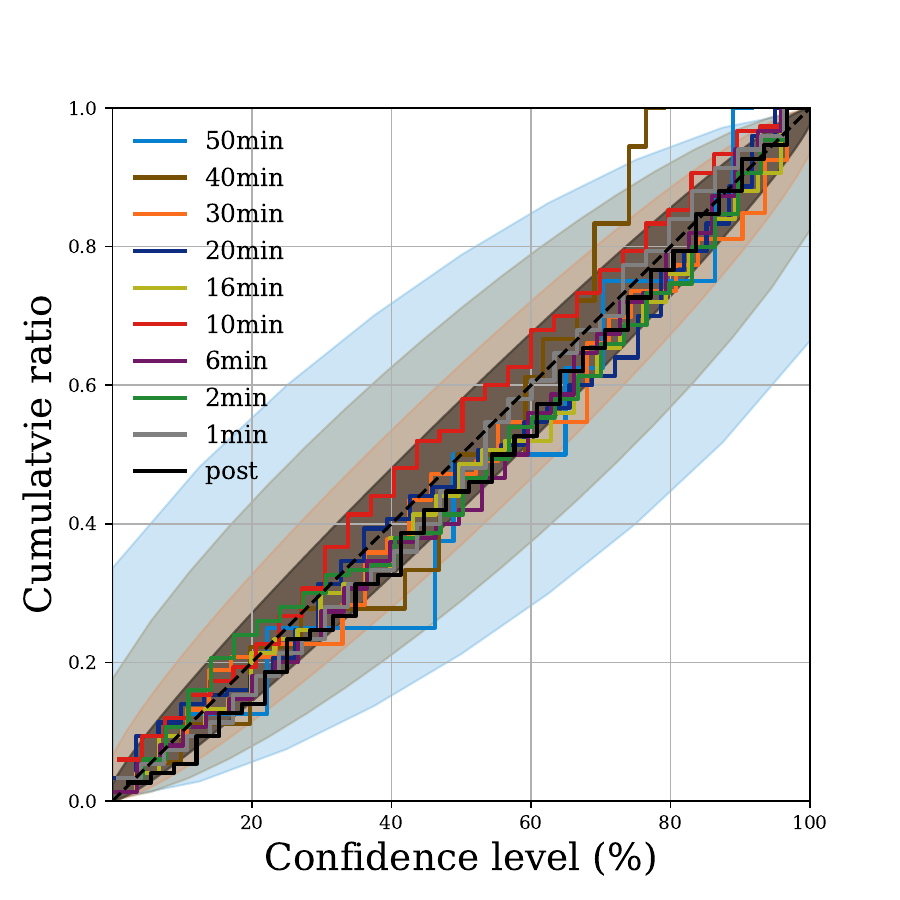}
    \centering
    \caption{\label{fig:ppplot} P-P plot of localizations in our simulation at different bands. {For 50\,min (light blue, sample size = 8), 40\,min (brown, sample size = 18), and 30\,min (orange, sample size = 53), error bars are plotted individually with their own colors. For other bands, we randomly select 150 events and plot their error bar in black. The error bar is calculated from a binomial distribution, and we note that it only converges to (0\%,~0) and (100\%,~1) when the sample size is sufficiently large.} }
\end{figure}

Fig.~\ref{fig:ppplot} is the P-P plot of our localization simulation, showing x\% confidence region (x-axis) is able to include y\% of total events (y-axis, scaled). The diagonal shapes shows the multi-band localization scheme is reasonably self-consistent. The lines for 30+ minutes before merger have larger statistical fluctuations due to the insufficient number of samples. 

\begin{figure}[b!]
    \includegraphics[width=0.46\textwidth]{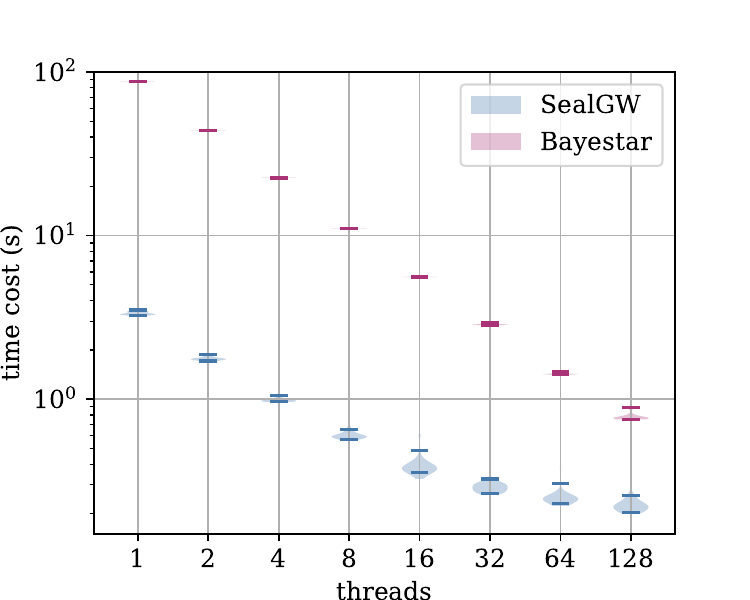}
    \centering
    \caption{\label{fig:timecost} Time cost of running \texttt{SealGW} {and \texttt{Bayestar}} for ET+2CE network on a 2.44 GHz processor with different number of threads, excluding the time costs of matched filtering and data conditioning. {The two algorithms are tested with the same data, and skymaps are calculated to the same level of resolution ($n_{side} = 2048$, finest pixel = 0.0008 deg$^2$). } The matched filtering speed (detection latency) depends on the efficiency of the detection pipeline, and data conditioning usually takes $\sim 0.1$s to read and pre-process the data from pipeline outputs.}
\end{figure}

We tested the time cost of \texttt{SealGW} {and \texttt{Bayestar} calculation with the same data (full-bandwidth ET+2CE network) and skymap resulution ($n_{side} = 2048$, finest pixel = 0.0008 deg$^2$), as showed in Fig.~\ref{fig:timecost}. Tests are performed on a 2.44 GHz processor with OpenMP multithreading. Thanks to the semi-analytical property, \texttt{SealGW} can achieve $\sim 26$ times faster speed than \texttt{Bayestar} with fewer threads, and the speed up factor goes down to $\sim 4$ when more threads come in as the non-parallelizable calculation begins to dominate \texttt{SealGW} run time. 
It only takes \texttt{SealGW} $\sim 3$s with 1 thread and $\sim 0.5$s with 8 threads, which means \texttt{SealGW} is able to perform real-time localization with a low hardware requirement. Note that the time cost can be further reduced with narrower bandwidth or coarser skymap resultions, e.g., time cost of \texttt{SealGW} can be halved when the finest resolution is 0.013deg$^2$.} The efficiency and cheapness is suitable for the 3G detector scenario in which number of detection can be huge. Nevertheless, a thorough estimate of early warning latency would require a comprehensive design of detection pipeline structure, and there would be a wall time of $\sim 0.1$s to read and preprocess the data from pipeline outputs. 

\section{\label{sec:conclusion} Conclusion and discussion}
We provide an exploratory demonstration of early warning localization of long signals for 3G GW detector networks. We simulate a mock catalog for one month of observation with an ET+2CE network, and perform multi-band analysis with the fast localization algorithm \texttt{SealGW}. We show that this is a efficient scheme for pre-merger localization. 

~

Multi-band analysis allows us to detect BNS in an early stage and update the results regularly with incoming data. There are tens of BNS detected more than 30 minutes before merger in our simulation, and localized within 100 deg$^2$ at $\sim 10$ minutes before merger. 10 deg$^2$ can be achieved $\sim 6$ minutes before merger. Since wide-field optical transient facilities usually have field of view of 1-10 deg$^2$ (see summaries in \citet{Sachdev:2020lfd, ronchini2022_PerspectivesMultimessengerAstronomy}), the precise pre-merger localization of BNS would be extremely helpful to finding EM counterparts before the merger and observing the entire process of BNS coalescence. 

{Our work here presents a solution for the crucial step of performing real-time localization in the context of online searches in 3G detectors, that effectively {reduces the latency and computational burden} arising from pre-merger localization. However there remains the larger issue of developing the surrounding infrastructure to search for pre-merger signals and disseminating sky-maps in low latency to observatories before detection.}
As an exploratory demonstration, we have made several simplifications to the problem, such as ignoring overlapping signals, assuming perfect matched filtering, and a relatively naive waveform segmentation. The merger rate estimation of BNS is also uncertain to date, therefore the absolute detection numbers should be interpreted as an order-of-magnitude estimation.
We plan to explore the multi-band analysis on a real detection pipeline 
and use a more accurate astrophysical population (which should be available in years with new observations) in our future work.

{\textit{Data availability}: The skymap fits files for 30, 20, 10, 6 and 1 minutes early warning and post-merger triggers are openly available in zenodo~\citep{data_release}. Further data and scripts for reproducibility are available upon reasonable request.}

\begin{acknowledgments}
We thank Linqing Wen, Daniel Tang and Chayan Chatterjee for helpful discussions. We are grateful for computational resources provided by Cardiff University, and funded by STFC grant ST/I006285/1. QH is supported by CSC. JV is supported by STFC grant ST/V005634/1.
\end{acknowledgments}

\bibliography{ewloc_refs.bib}
\bibliographystyle{aasjournal}

\end{document}